# Beamforming Matrix Quantization with Variable Feedback Rate

Chau Yuen, Sumei Sun
Institute for Infocomm Research (I²R), Singapore
{cyuen, sunsm}@i2r.a-star.edu.sg

Mel Meau Shin Ho
Imperial College, UK
m.ho@imperial.ac.uk

*Abstract:* We propose an improved beamforming matrix compression by Givens Rotation with the use of variable feedback rate. The variable feedback rate means that the number of bits used to represent the quantized beamforming matrix is based on the value of the matrix. Compared with the fixed feedback rate scheme, the proposed method has better performance without additional feedback bandwidth.

## I. INTRODUCTION

Multiple transmit and receive antennas system has been considered in several communication standards in order to achieve a higher throughput. Although open-loop multiple-input multiple-output (MIMO) technique has already shown to achieve high performance gain, the availability of either full or partial channel state information (CSI) at the base-station will still lead to additional performance gain and complexity reduction. Such closed-loop schemes have been considered in many communication standards for application of beamforming or multi-user precoding.

However, CSI estimation for the downlink channel at the base-station is not possible in FDD systems, it is also not straight forward in TDD systems due to the mismatch in the radio front end. Hence in general, the CSI will be estimated at the mobile and sent back to the base-station. This, unfortunately, requires a high feedback bandwidth. So the mobile may compute the beamforming matrix, which is usually a unitary matrix, and "compress" such matrix before feeding back to the base-station. The "compression" can greatly reduce the feedback bandwidth requirement.

There have been several proposals in the literature to compress the beamforming vector. One is codebook based such as vector quantization proposed in [2][3], and another is by using Givens Rotation (GR) [4]. Compared with the GR-based scheme, the codebook-based approach requires a higher storage, as a set of codebook is needed for a particular antenna setting. It has a higher complexity than GR, especially when the number of codewords in the codebook increases. Due to these reasons, GR has been adopted in 802.11n standard [4].

In this paper, we investigate an effective way to compress the beamforming matrix with GR, such that our proposed scheme can achieve a better performance than existing technique without causing extra bandwidth.

## II. SIGNAL MODEL

### A) MIMO Model

Consider a point-to-point MIMO channel with $N_T$ transmit antennas and $N_R$ receive antennas, the $N_T \times 1$ transmitted signal denoted as $\mathbf{x}$ and the $N_R \times N_T$ channel denoted as $\mathbf{H}$. The $N_R \times 1$ received signal $\mathbf{y}$ can be expressed as:

$$\mathbf{y} = \mathbf{Hx} + \mathbf{n} \qquad (1)$$

To demonstrate the idea of beamforming, we use eigen-subspace beamforming as an example. By using Singular Value Decomposition (SVD), a MIMO channel $\mathbf{H}$ can be decomposed into:

$$\mathbf{H} = \mathbf{UDV}^H \qquad (2)$$

where $\mathbf{U}$ of size $N_R \times R$ and $\mathbf{V}$ of size $N_T \times R$ are both unitary matrices, and $\mathbf{D}$ is a $R \times R$ diagonal matrix consisting of the singular value of $\mathbf{H}$ as its diagonal elements, and $R$ is the rank of $\mathbf{H}$. To perform eigen-subspace beamforming, $\mathbf{V}$ needs to be fedback to the base-station. In order to reduce the amount of information in $\mathbf{V}$, it is proposed in [1] that we should multiply $\mathbf{V}$ with a matrix $\boldsymbol{\Sigma}$ such that the last row of $\mathbf{V}$ consists of only real numbers. Hence we may re-express (2) as:

$$\mathbf{H} = \mathbf{UD\Sigma}\overline{\mathbf{V}}^H = \mathbf{U}\overline{\mathbf{D}}\overline{\mathbf{V}}^H \quad \text{where } \overline{\mathbf{D}} = \mathbf{D\Sigma} \qquad (3)$$

where

$$\boldsymbol{\Sigma} = \text{diag}\left[\exp\left\{j*\arg\left(\overline{\mathbf{v}}_{N_T}^H\right)\right\}\right] \qquad (4)$$

and $\overline{\mathbf{v}}_{N_T}^H$ represents the last column of $\overline{\mathbf{V}}^H$.

$$\mathbf{y} = \mathbf{Hx} + \mathbf{n} = \left(\mathbf{U}\overline{\mathbf{D}}\overline{\mathbf{V}}^H\right)\mathbf{x} + \mathbf{n} \qquad (5)$$

To transmit data on the first $K$ eigen modes (where $K \leq R$), the beamforming matrix $\mathbf{W}$ is the first $K$ column vectors of $\overline{\mathbf{V}}$:

$$\mathbf{W} = \overline{\mathbf{V}}_{(1:K)} \qquad (6)$$

The transmitted signal is related to the $K \times 1$ data signal $\mathbf{u}$ by:

$$\mathbf{x} = \mathbf{Wu} \qquad (7)$$

To retrieve data, the mobile multiplies the received signal with $\mathbf{U}^H$,

$$\hat{\mathbf{u}} = \mathbf{U}^H \mathbf{y} = \overline{\mathbf{D}}_{(1:K)} \mathbf{u} + \tilde{\mathbf{n}} \qquad (8)$$

where $\tilde{\mathbf{n}}$ has the same statistics as $\mathbf{n}$ (as $\mathbf{U}$ is a unitary matrix). And $\overline{\mathbf{D}}$ is a diagonal matrix, hence eigen-beamforming leads to simple decoding, as the MIMO channel can be treated as a few parallel subchannels.

In practice, due to the limited bandwidth in the feedback channel, $\mathbf{W}$ has to be quantized, and the base-station receives the quantized version of $\mathbf{W}$, denoted as $\tilde{\mathbf{W}}$. We assume the channels are estimated accurately, and there is no error or delay in the feedback channel, so we only consider the impact of quantization error due to limited feedback bandwidth. Hence $\tilde{\mathbf{W}}$, instead of $\mathbf{W}$, is being used as beamforming matrix. In this paper, we are going to present an effective way to quantize $\mathbf{W}$ such that the quantization error is smaller than existing methods by using the same average number of feedback bits.

In this paper, we use only eigen-beamforming based on SVD as an example, however it does not restrict the application of our proposed method to other types of beamformer or applications, such as multi-user MIMO precoding.

**B) Givens Rotation Model**

As our new idea can be easily applied on GR to improve its performance, we give a brief review on GR. A unitary matrix, such as **W** in our case, can be represented as follows:

$$\mathbf{W} = \prod_{i=1}^{\min(N_T-1,K)} \left[ \mathbf{D}_i \begin{pmatrix} 1_{i-1} & e^{j\phi_{i,i}} & \cdots & e^{j\phi_{N-1,i}} \end{pmatrix} \prod_{l=i+1}^{N_T} \mathbf{G}_{li}^T(\psi_{li}) \right] \times \mathbf{I}_{N_T \times K} \quad (9)$$

where $\mathbf{D}_i$ is a diagonal matrix and **G** is defined as:

$$\mathbf{G}_{li}(\psi_{li}) = \begin{bmatrix} \mathbf{I}_{i-1} & 0 & 0 & 0 & 0 \\ 0 & \cos(\psi_{li}) & 0 & \sin(\psi_{li}) & 0 \\ 0 & 0 & \mathbf{I}_{l-i-1} & 0 & 0 \\ 0 & -\sin(\psi_{li}) & 0 & \cos(\psi_{li}) & 0 \\ 0 & 0 & 0 & 0 & \mathbf{I}_{N_T-1} \end{bmatrix} \quad (10)$$

Take a $3 \times 2$ unitary matrices **W** as example, it can be described as:

$$\mathbf{W} = \begin{bmatrix} e^{j\phi_{11}} & 0 & 0 \\ 0 & e^{j\phi_{21}} & 0 \\ 0 & 0 & 1 \end{bmatrix} \times \mathbf{G}_{21}^T(\psi_{21}) \mathbf{G}_{31}^T(\psi_{31}) \times \\ \begin{bmatrix} 1 & 0 & 0 \\ 0 & e^{j\phi_{22}} & 0 \\ 0 & 0 & 1 \end{bmatrix} \times \mathbf{G}_{32}^T(\psi_{32}) \times \begin{bmatrix} 1 & 0 \\ 0 & 1 \\ 0 & 0 \end{bmatrix} \quad (11)$$

Hence the $3 \times 2$ unitary matrices **W** can be fully described by just six parameters: $\phi_{11}$, $\phi_{21}$, $\psi_{21}$, $\psi_{31}$, $\phi_{22}$, $\psi_{32}$.

Table 1 lists down the number of parameters for GR of various dimensions.

Table 1 Number of parameters for GR

| Dimension of W | Number of parameters | Parameters |
|---|---|---|
| $2 \times 1$, $2 \times 2$ | 2 | $\phi_{11}$, $\psi_{21}$ |
| $3 \times 1$ | 4 | $\phi_{11}$, $\phi_{21}$, $\psi_{21}$, $\psi_{31}$ |
| $3 \times 2$, $3 \times 3$ | 6 | $\phi_{11}$, $\phi_{21}$, $\psi_{21}$, $\psi_{31}$, $\phi_{22}$, $\psi_{32}$ |
| $4 \times 1$ | 6 | $\phi_{11}$, $\phi_{21}$, $\phi_{31}$, $\psi_{21}$, $\psi_{31}$, $\psi_{41}$ |
| $4 \times 2$ | 10 | $\phi_{11}$, $\phi_{21}$, $\phi_{31}$, $\psi_{21}$, $\psi_{31}$, $\psi_{41}$, $\phi_{22}$, $\phi_{32}$, $\psi_{32}$, $\psi_{42}$ |
| $4 \times 3$, $4 \times 4$ | 12 | $\phi_{11}$, $\phi_{21}$, $\phi_{31}$, $\psi_{21}$, $\psi_{31}$, $\psi_{41}$, $\phi_{22}$, $\phi_{32}$, $\psi_{32}$, $\psi_{42}$, $\phi_{33}$, $\psi_{43}$ |

The number of bits assigned to the GR parameters in IEEE 802.11n draft is summarized in

Table 2, where $b_\psi$ represents the number of bits assigned to $\psi$, and $b_\phi$ represents the number of bits assigned to $\phi$. The unequal bit assignment is due to different range of $\psi$ and $\phi$, $\psi$ has a range from 0 to $\pi/2$, while $\phi$ has a range from 0 to $2\pi$ [4].

Table 2 Bits assignment for GR parameters

| $b_\psi$ | 1 | 2 | 3 | 4 |
|---|---|---|---|---|
| $b_\phi$ | 3 | 4 | 5 | 6 |

Using the bit assignment in

Table 2, $\psi$ and $\phi$ can be quantized according to (12) (where $\tilde{\psi}$ and $\tilde{\phi}$ represent the quantized version of $\psi$ and $\phi$), the beamforming matrix $\tilde{\mathbf{W}}$ can be recovered at the base-station by using (14).

$$\tilde{\psi} = \frac{k\pi}{2^{b_\psi+1}} + \frac{\pi}{2^{b_\psi+2}} \quad \text{where } k = 0,1,\ldots,2^{b_\psi} - 1 \quad (12)$$

$$\tilde{\phi} = \frac{k\pi}{2^{b_\phi-1}} + \frac{\pi}{2^{b_\phi}} \quad \text{where } k = 0,1,\ldots,2^{b_\phi} - 1 \quad (13)$$

$$\tilde{\mathbf{W}} = \prod_{i=1}^{\min(N_T-1,K)} \left[ \mathbf{D}_i \begin{pmatrix} 1_{i-1} & e^{j\tilde{\phi}_{i,i}} & \cdots & e^{j\tilde{\phi}_{N_T-1,i}} \end{pmatrix} \prod_{l=i+1}^{N_T} \mathbf{G}_{li}^T(\tilde{\psi}_{li}) \right] \times \mathbf{I}_{N_T \times K} \quad (14)$$

## III. PROPOSED SCHEME

The three basic ideas of the proposed scheme are as follows:
1) *Dynamic Bit Assignment*: The bits assigned to Givens Rotation parameters $\phi$ can be depended on the value of parameters $\psi$. So when the resolution is "sparse", we use more bits for the quantization of $\phi$; when the resolution is "crowded", we use fewer bits for $\phi$. In other words, the bit assignment to $\phi$ is adaptively adjusted based on the value of $\psi$.
2) *Efficient Source Coding*: Due to the non-uniform distribution of the Givens Rotation parameters $\psi$, efficient source coding, such as Huffman code [5], can be used to efficiently encode the Givens Rotation parameters $\psi$ and hence reduce the number of feedback bits required.
3) *Codebook design*: Instead of quantizing the Givens Rotation parameters in a uniform manner, codebook can be designed for the Givens rotation parameters $\phi$ and $\psi$.

Each of the above three ideas can be combined together or applied separately.

**A) Dynamic Bit Assignment**

To illustrate the idea, it is best to make use of a simple example of $2 \times 1$ beamforming vector. Consider a $2 \times 1$ unit-norm vector **w** as shown in (15), due to unit norm property, it must satisfy the constraints in (16). In addition, there is a matching between *Givens Rotation* view point with the *Geometry* view point, i.e. $r_1$ and $r_2$ are related to $\psi_{21}$ ($r_1 = \cos(\psi_{21})$ and $r_2 = \sin(\psi_{21})$).

$$\mathbf{w} = \begin{bmatrix} w_1 \\ w_2 \end{bmatrix}$$
$$= \begin{bmatrix} r_1 e^{j\phi_{11}} \\ r_2 \end{bmatrix} \leftarrow \text{Geometry view point} \quad (15)$$
$$= \begin{bmatrix} \cos\psi_{21} e^{j\phi_{11}} \\ \sin\psi_{21} \end{bmatrix} \leftarrow \text{Givens Rotation view point}$$

$$w_1^2 + w_2^2 = 1 \Rightarrow r_1^2 + r_2^2 = 1 \quad (16)$$

Based on the geometry view point in (15) and the constraints in (16), we obtain the following insights:
- When $r_2$ is large ($\psi_{21}$ is large), $r_1$ will be small; $\phi_{11}$ can have a lower resolution.
- When $r_2$ is small ($\psi_{21}$ is small), $r_1$ will be large; $\phi_{11}$ will need a higher resolution.

Hence these suggest us that, the bit assigned to $\phi_{11}$ should be a function of $r_1$ and $r_2$, which is in turn related to the value of $\psi_{21}$ when GR is in use. To have a better idea, we illustrate our idea using Figure 1 and Figure 2. As shown in Figure 1, the radii of two circles represent two possible values of $r_1$ (this is equivalent to one bit assignment to $\psi_{21}$). When $r_2$ is small, $r_1$ will be large (this corresponds to the blue circle marked with x at the bottom of Figure 1); when $r_2$ is large, $r_1$ will be small (this corresponds to the red circle marked with o at the top of Figure 1). It can be seen that in this case, if we assign same number of bits (e.g. 3 bits) to $\phi_{11}$, it corresponds to eight points on each circle. And the points on the upper circle are closer to each other, while the points on the lower circle are further apart. In this case, the total number of bits to represent **w** is 1+3 = 4 bits.

To achieve a lower quantization error, we can assign different number of bits to $\phi_{11}$ according to the value of $r_1$. For example, as shown in Figure 2, we assign two bits for $\phi_{11}$ when $r_1$ is small (i.e. $\psi_{21}$ is large, the upper circle), and we assign four bits for $\phi_{11}$ when $r_1$ is large (i.e. $\psi_{21}$ is small, the lower circle). It can be seen that the distance between the points are more evenly distributed in this case. Depending on the assignment of $\psi_{21}$, we can have the probability of having two cases (upper or lower circle) to be equal. Hence in this case, the total number of bits to represent **w** is 1+2=3 or 1+4=5, which is 4 bits on average.

An optimal codebook obtained by vector quantization methodology as described in [2] is shown in Figure 3. It can be seen that by using variable feedback rate, the final codebook appears closer to the optimal codebook than the one with fixed feedback rate.

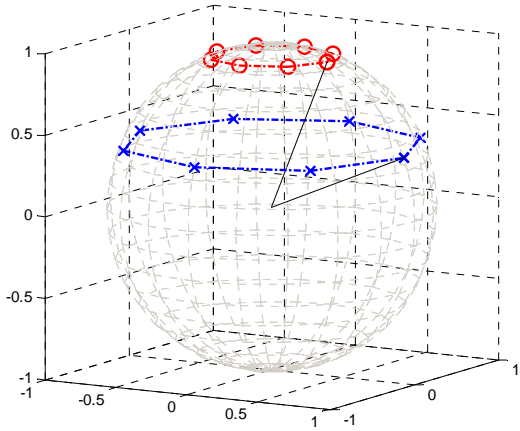

Figure 1: Distribution of quantized 2-by-1 beamforming vector based on fixed feedback rate

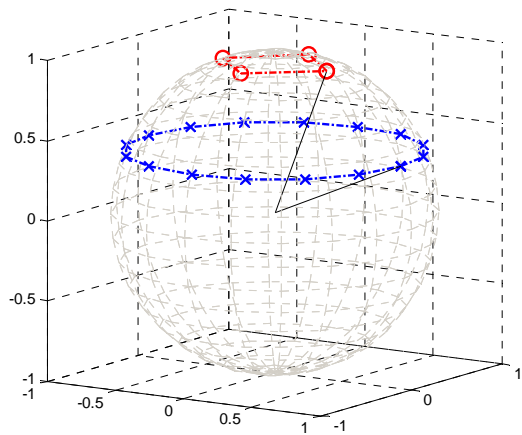

Figure 2: Distribution of quantized 2-by-1 beamforming vector based on variable feedback rate

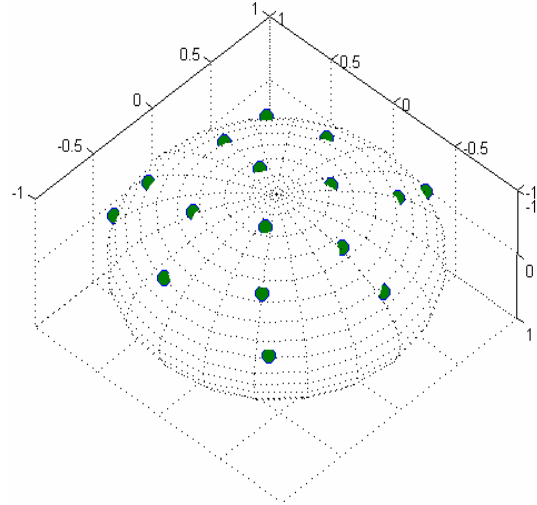

Figure 3: Optimal codebook trained by vector quantization

## B) Efficient Source Coding

In Figure 4(a) to (f), we show the distribution of the GR parameters for a unitary beamforming matrix of dimension 3x2. It can be seen that the parameters $\phi$ have a uniform distribution from 0 to 360 degrees (i.e. 0 to $2\pi$), while the parameters $\psi$ have non-uniform distribution from the range of 0 to 90 degree (i.e. 0 to $\pi/2$), an asymmetric distribution for $\psi_{31}$ and a symmetric distribution for $\psi_{21}$ and $\psi_{32}$. We also show the distribution of the quantized version of the parameters $\psi$ when using 4 levels of granularity in Figure 4(g) to (i).

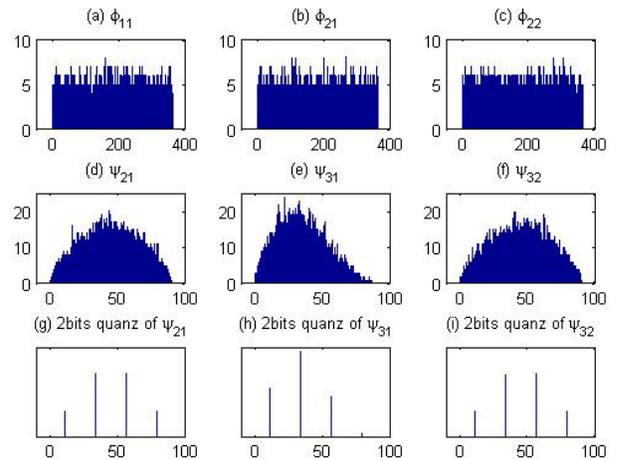

Figure 4: Distribution of Givens Rotation parameters for 3-by-2 beamforming matrix

## IV. CASE STUDY

Depending on the training symbol placement and receivers' design, we consider two cases. In the first case, a simple receiver is not retrained with the beamforming matrix, hence the receiver do not take the mismatch of the quantized beamforming matrix into account, and simply use a parallel decoder. In the second case, the receiver is retrained with the updated beamforming matrix, hence the mismatch between the quantized beamforming matrix and the channel is taken into account, and use a more complicated receiver, such as MMSE receiver .

### A) Receiver with simple parallel decoder

In this section, we consider the receiver as stated in (8), which is repeated as below by taking into account the mismatch in the beamforming matrix with the channel:

$$\hat{\mathbf{u}} = \mathbf{U}^H \mathbf{y} = \bar{\mathbf{D}}\bar{\mathbf{V}}\tilde{\mathbf{W}}\mathbf{u} + \tilde{\mathbf{n}} \quad (17)$$

Due to quantization, $\bar{\mathbf{V}}\tilde{\mathbf{W}}$ is no longer an identity matrix, hence such a simple receiver would be highly sensitive to the quantization error.

In this section, we demonstrate in details on how the proposed scheme works for 3 × 1 beamforming vector, which is as shown in (18). In Table 3 we show the bit assignment for the GR of the traditional and the newly proposed scheme. For example, for average 8 bits feedback, the traditional scheme assigns 1 bits for $\psi_{21}$ and $\psi_{31}$, and 3 bits for $\phi_{11}$ and $\phi_{21}$, while in the proposed scheme, 1 bit is assigned for $\psi_{21}$ and $\psi_{31}$, but 2, 3 or 4 bits for $\phi_{11}$, with the actual assignment (whether 2 bits, 3 bits or 4 bits) depending on the value of $\psi_{21}$ and $\psi_{31}$, which is shown in Table 4.

It should be noted that we have designed a codebook for the value of $\psi$ (as specified below in Table 4).

$$\mathbf{w} = \begin{bmatrix} r_1 e^{j\phi_1} \\ r_2 e^{j\phi_{21}} \\ r_3 \end{bmatrix} = \begin{bmatrix} \cos\psi_{21}\cos\psi_{31}e^{j\phi_{11}} \\ \sin\psi_{21}\cos\psi_{31}e^{j\phi_{21}} \\ \sin\psi_{31} \end{bmatrix} \quad (18)$$

Table 3 Number of bits allocation for three transmit antennas beamforming vector with ave 8 bits feedback

| | $b_{\psi_{31}}$ | $b_{\psi_{21}}$ | $b_{\phi_{21}}$ | $b_{\phi_{11}}$ | Remark |
|---|---|---|---|---|---|
| Traditional | 1 | 1 | 3 | 3 | |
| Proposed | 1 | 1 | 2, 3, 4 | 2, 3, 4 | Table 4 |

*Remark: the assignment of number of bits is just an example, there could be other assignments that lead to better performance.*

Table 4 Bit allocation for $\phi_{21}$ and $\phi_{11}$ when ave. 8 bits feedback

| Bit representative of | | Bits allocated for $\phi_{21}$ and $\phi_{11}$ | | Total number of feedback bits |
|---|---|---|---|---|
| $\psi_{31}{}^*$ | $\psi_{21}{}^*$ | $b_{\phi_{21}}$ | $b_{\phi_{11}}$ | |
| 0 | 0 | 3 | 4 | 9 |
| 0 | 1 | 4 | 3 | 9 |
| 1 | 0 | 2 | 3 | 7 |
| 1 | 1 | 3 | 2 | 7 |

*Possible values for $\psi_{31}$ and $\psi_{21}$ are [0.2967 0.8727] rad.

We first compare the error in quantization by using mean square error (MSE) in (19) or mean angular distance (MAD) [3] in (20).

$$MSE = \mathrm{E}|\mathbf{w} - \tilde{\mathbf{w}}|^2 \quad (19)$$

$$MAD = \mathrm{E}\sqrt{1 - |\mathbf{w} \cdot \tilde{\mathbf{w}}|^2} \quad (20)$$

where • in (20) represents dot product.

Table 5 shows the MSE and MAD for quantization of 3 × 1 beamforming vector based on traditional fixed rate feedback and the newly proposed scheme based on variable rate feedback. We show that the newly proposed scheme achieves a lower MSE and MAD than the traditional scheme with average 8 bits feedback.

Table 5 MSE and MAD comparisons

| | MSE | MAD |
|---|---|---|
| Traditional | 0.110 | 0.312 |
| Proposed | 0.092 | 0.282 |

The BER performance is shown in Figure 5. It can be seen that the proposed scheme outperforms the traditional way. And such performance gain is achieved without any additional feedback bandwidth.

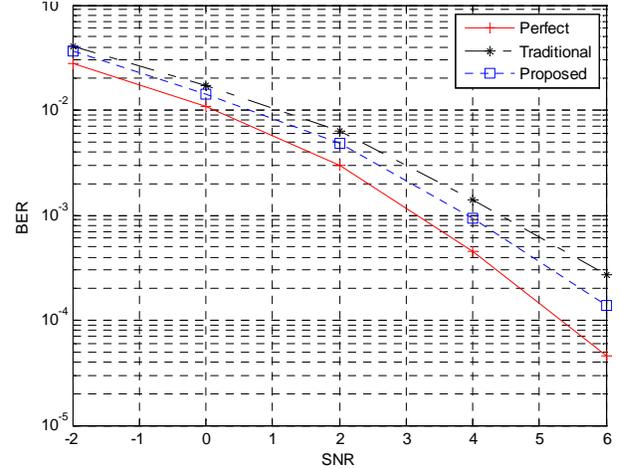

Figure 5: 3 tx 3 rx, one stream based on SVD beamforming, QPSK, average 8 bits feedback

### B) Receiver with simple parallel decoder

In this section, we consider a different receiver from that in (8). It is assumed that the mismatch in the beamforming matrix is known, and apply the MMSE detector as shown below:

$$\hat{\mathbf{u}} = \left(\mathbf{G}^H\mathbf{G} + \alpha\mathbf{I}\right)^{-1}\mathbf{G}^H\mathbf{y} \quad (21)$$

$$\mathbf{G} = \bar{\mathbf{D}}\bar{\mathbf{V}}\tilde{\mathbf{W}}$$

In this case, we consider, a three transmit and three receive antennas system, and involves two streams of data. Due to the non-uniform distribution of the parameters $\psi$ as shown in Figure 4, we can make use of the Huffman code [2] to encode the quantized value of $\psi$. As shown in Table 6 and Table 7, we shows that it is enough to represent the quantized version of $\psi_{21}$ (or $\psi_{32}$) and $\psi_{31}$ by just using 1.94 and 1.77 bits (instead of 2 bits for a granularity of four).

Table 6 Huffman code for GR parameters $\psi_{21}$ and $\psi_{32}$

| Quantized value of $\psi_{21}$ or $\psi_{32}$ | Probability (i) | Huffman Code | Bits for $\psi_{21}$ or $\psi_{32}$, (ii) | Ave bits for $\psi_{21}$ or $\psi_{32}$, (i)*(ii) |
|---|---|---|---|---|
| 11.25 | 0.14714 | 110 | 3 | 0.44142 |
| 33.75 | 0.35496 | 0 | 1 | 0.35496 |
| 56.25 | 0.35146 | 10 | 2 | 0.70292 |
| 78.75 | 0.14644 | 111 | 3 | 0.43932 |
| | | | | ***1.93862*** |

Table 7 Huffman code for GR parameters $\psi_{31}$

| Quantized value of $\psi_{31}$ | Probability (i) | Huffman Code | Bits for $\psi_{31}$, (ii) | Ave bits for $\psi_{31}$, (i)*(ii) |
|---|---|---|---|---|
| 11.25 | 0.2722 | 10 | 2 | 0.5444 |
| 33.75 | 0.47748 | 0 | 1 | 0.47748 |
| 56.25 | 0.2299 | 110 | 3 | 0.6897 |
| 78.75 | 0.02042 | 111 | 3 | 0.06126 |
| | | | | *1.77284* |

It is also found that when $\psi_{21}$= 33.75, 56.25 and $\psi_{31}$ = 11.25, 33.75, the quantized version of the beamforming matrix will have higher chances of being poor quality, hence we suggest to use 3 bits to quantize $\phi_{11}$ and $\phi_{21}$ when the above-mentioned condition happens. The detailed bit assignment for the $\phi_{11}$ and $\phi_{21}$ is omitted in this paper.

Combining all the above, the average number of bits required to represent a 3x2 unitary beamforming matrix using our proposed scheme can be computed as 12.7 bits (detailed computation is omitted in this paper).

We compare five different quantization schemes in Table 8. It can be seen that our proposed scheme (Scheme E) achieve a better performance than Scheme B and C with 0.71 bits higher in average feedback bits, while Scheme D achieves an even better performance but with 3 bits higher in average feedback bits than Schemes B and C.

Table 8 Five schemes for comparisons

| | Bit assignment | Ave number of feedback bits |
|---|---|---|
| Scheme A | Perfect feedback | ∞ |
| Scheme B | $b_\psi$ = 1 and $b_\phi$ = 3 | 12 |
| Scheme C | $b_\psi$ = 2 and $b_\phi$ = 2 | 12 |
| Scheme D | $b_\psi$ = 2 and $b_\phi$ = 3 | 15 |
| Scheme E (proposed) | Table 6, Table 7 | 12.71 |

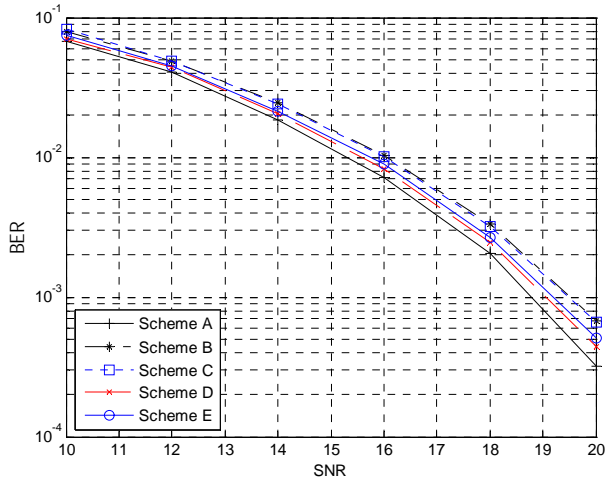

Figure 6  BER performance of the first stream

The simulation results, BER of the five schemes are shown in Figure 6. We assume three transmit antennas and three receive antennas with eigen beamforming, two data streams: one 64QAM and the other 16QAM, hence a 10 bits/sec/Hz spectral efficiency is achieved. It can be seen that Schemes B and C perform the worst, while Scheme A performs the best. Scheme D can perform better than Schemes B and C by using 3 bits more feedback bandwidth. Our proposed scheme has a performance close to Scheme D, but we only require 0.71 bits more feedback bandwidth than Schemes B and C.

As shown by the case study in this paper, by using merely extra 0.7 bits, we can obtain half of the gain that is obtained with 3 extra bits. By further investigation, it is believed the proposed scheme can offer significant performance gain.

The newly proposed scheme can be considered as a hybrid of traditional GR approach and codebook based approach, i.e. we have a code book for the GR parameters $\phi$ and $\psi$. However the new scheme has a lower storage than those based on codebook case.

## V. Conclusion

In this paper, a simple quantization scheme has been presented for unit-norm beamforming vector based on variable feedback rate. The basic idea is to give higher resolution in the dense area and lower resolution in the other case. The idea can be directly applied on existing Givens Rotation (GR), by having the bits allocated to $\phi$ parameters to be dependent on the value of $\psi$ parameters. Due to non-uniform distribution of GR parameter $\psi$, the performance can be further improved if we consider efficient source coding and codebook design for GR parameters. Results show that the newly proposed scheme can achieve a lower mean square error and lower mean angular distance. The BER performance of the close-loop MIMO system based on the proposed quantization scheme also outperforms existing schemes.

The proposed idea is not restricted to the use of eigen-beamformer or GR, which have been used to demonstrate the idea in the paper. Our proposed method gives a better accuracy when compressing a unitary matrix, and such accuracy plays an important role in many communications system, that include precoding for multi-user MIMO.

## References

[1] J. Kim and C. Aldana, "Efficient feedback of the channel information for closeloop beamforming in WLAN", *IEEE VTC-Spring 2006*.
[2] J. C. Roh and B. D. Rao, "Channel feedback quantization methods for MISO and MIMO systems", *IEEE PIMRC 2004*, pp. 805-809.
[3] D. J. Love, R. W. Heath, and T. Strohmer, "Grassmannian beamforming for multiple-input multiple-output wireless system", *IEEE Trans. on Information Theory*, vol. 49, pp. 2735-2747, Oct 2003.
[4] IEEE P802.11n/D001, IEEE Standards Draft, Jan 2006.
[5] T. M. Cover and J. A. Thomas, "Elements of information theory", Wiley Interscience, 1991.